\documentclass[useAMS,usenatbib]{mn2e}
\usepackage{graphicx}
\usepackage[section]{placeins}

\usepackage{times} 
\usepackage{amsmath}
\usepackage{amssymb}
\usepackage{tikz}
\usepackage{xcolor}

\begin{document}

\title[Resonance capture at arbitrary inclination]{Resonance capture at arbitrary inclination: II. Effect of the radial drift rate}
\author[F. Namouni and M. H. M. Morais]{F. Namouni$^{1}$\thanks{E-mail:
namouni@obs-nice.fr (FN) ; helena.morais@rc.unesp.br (MHMM)} and  M. H. M. Morais
$^{2}$\footnotemark[1]\\
$^{1}$Universit\'e C\^ote d'Azur, CNRS, Observatoire de la C\^ote d'Azur, CS 34229, 06304 Nice, France\\
$^{2}$Instituto de Geoci\^encias e Ci\^encias Exatas, Universidade Estadual Paulista (UNESP), Av. 24-A, 1515 13506-900 Rio Claro, SP, Brazil}

\date{Accepted 2017 January 31. Received 2017 January 30; in original form 2016 August 25.}

\maketitle

\begin{abstract}
The effect of radial drift rate on  mean motion resonance capture is studied for prograde, polar and retrograde orbits. We employ the numerical framework of our earlier exploration of resonance capture at arbitrary inclination. Randomly constructed samples of massless particles are set to migrate radially from outside the orbit of a Jupiter-mass planet at different drift rates totalling more than $1.6\times 10^6$ numerical simulations. Slower drift rates reduce overall capture probability especially for prograde orbits and enhance capture at specific initial inclinations of high order resonances such as the outer 1:5, 1:4, 1:3, 2:5, 3:7, 5:7. Global capture is reduced with increasing eccentricity at all inclinations  as high order resonances capture more particles that are subsequently lost by disruptive close encounters with the planet. The relative efficiency of retrograde resonances at long-lived capture with respect to prograde resonances is explained by the reduced effect of planet encounters as such events occur with a shorter duration and a higher relative velocity for retrograde motion.  Capture in the coorbital 1:1 resonance  is marginally affected by the radial drift rate except for nearly coplanar retrograde eccentric orbits whose capture likelihood is increased significantly with slower drift rates. An unexpected finding is the presence of a dynamical corridor for capture in  high order inner prograde resonances  with initial inclinations in the range [50$^\circ$,80$^\circ$] especially at the inner 5:2 resonance whose capture likelihood peaks at 80\% to 90\% depending on initial eccentricity.

\end{abstract}


\section{Introduction}
The presence of Centaurs and Damocloids (2006 BZ8, 2008 SO218, 2009 QY6 and 1999 LE31)  in retrograde resonance  with Jupiter and Saturn  \citep{MoraisNamouni13b} has recently instigated research into the mechanisms of resonance capture at arbitrary inclination. {Retrograde resonance refers to mean motion resonance with the minor body on retrograde motion or equivalently with an inclination larger than $90^\circ$.} The most unexpected {recent} result was that retrograde resonances are more efficient at orbital capture than prograde resonances (\cite{NamouniMorais15} hereafter Paper I). For the Centaurs inherently chaotic orbits that wander radially between the planets to be captured then released from mean motion resonances \citep{BaileyMalhotra09,VolkMalhotra13}, the efficiency of retrograde resonance capture suggests that the dynamical lifetimes of Centaurs and Damocloids on retrograde orbits are probably the largest of their kind in the solar system. 

This conclusion about the efficiency of retrograde resonances was established in Paper I using large scale simulations (nearly $6\times10^5$) by setting up massless particles that interact with a Jupiter mass planet while they drift radially under the effect of a Stokes-type friction force with a characteristic timescale of $10^5$ planetary orbits ($1.2\times 10^6$ years). Whereas this timescale is much larger than the involved planetary periods making radial drift an adiabatic orbital process, smaller drift rates are known to boost capture likelihood for planar higher order resonances. Using the analytical  Poincar\'e model of resonance in the planar three-body problem \citep{Poincare02,Henrard1,Lemaitre84,BorderiesGoldreich84,Henrard2,Henrard3,Henrard4,ssdbook}, \cite{Quillen06} examined how capture probability in first and second order resonances  may be reduced {through various parameters such as  semimajor axis drift times and initial particle and planet eccentricities.} It was found that probability capture tends to be reduced with shorter drift times, larger particle eccentricity and a finite orbital eccentricity of the planet. 

In order to ascertain the effect of longer radial drift times on resonant capture at arbitrary inclination, we may not use analytical tools such as the Poincar\'e model of resonance  {\citep{MoraisGiuppone12}} as we explained in detail in Paper I. At resonance passage or capture, eccentricity and inclination growth may be significant and three-body dynamics becomes complex:  the activation of higher harmonics near the nominal resonant location may or may not result in resonance overlap {\citep{Wisdom80,Wisdom82,Wisdom83}} but it does influence the capture of the drifting particle by temporarily trapping it in high order resonances before its arrival to the main low order resonance. In addition, the secular evolution of eccentricity and inclination through the Kozai-Lidov  mechanism  {\citep{Kozai62,Lidov62,GronchiMilani99, LithwickNaoz11,Katzetal11}} makes resonance crossing more complex through the three stages of the capture process uncovered in Paper I {(see section 3.1)}. These facts  had motivated one of our main conclusions that resonance order is not necessarily a good indicator of capture efficiency at arbitrary inclination. 

For these reasons, we revisit the numerical set-up of Paper I and carry out more simulations with longer drift times in order to measure the capture efficiency of the main thirty resonances associated with the Jupiter-mass planet for various eccentricity distributions. Particle inclination is varied between $0^\circ$ and $180^\circ$ with a $2.5^\circ$ step for the main simulations improving on the $5^\circ$ initial inclination resolution of Paper I. In section 2, we discuss the numerical setup and initial conditions. In section 3, we report the measured capture likelihoods and how they compare to those of Paper I. Section 4 contains our conclusions.

 \section{Numerical setup} 
In this investigation, we use the three-body problem of Paper I consisting of a solar-mass { and solar-radius} star, a Jupiter-mass { and Jupiter-radius} planet on a circular orbit with { semi-major axis $a_p$} and a massless particle with { semi-major axis $a_0=3a_p$} outside the 1:5 resonance with the planet (located at semi-major axis $2.924 a_p$). An imposed slow inward semi-major axis drift will make the particle's orbit sweep all  major resonances. {Drift is modelled through a velocity-dependent drag force of the form $-k {\bf v}$ leading to { a semi-major axis function $a(t)=a_0\exp(-2k t)$. } This drag force does not model a specific physical process that the particles may be subjected to, it is used to study resonance crossing with the simplest deterministic and unidirectional radial drift that does influence only the semi-major axis evolution and leaves the other orbital elements such as eccentricity and inclination unchanged. T}hree main characteristic drift times are used $\tau=(2k)^{-1}=10^{5}, \ 10^6$ and $10^7$ planetary orbital periods (in terms of physical units: $1.2 \times 10^{6}, 1.2 \times  10^7$ and $1.2 \times 10^8$ years) --unless stated otherwise $\tau$ is quoted in planetary periods. {In Paper I, only the drift rate $\tau=(2k)^{-1}=10^{5}$ was used.}
The effect of the planet's eccentricity was examined in Paper I. Here we shall maintain the planet on a circular orbit. The particle's inclination is taken from the following three sets [2.5$^\circ$:177.5$^\circ$] with a 2.5$^\circ$-step, [0.01$^\circ$:1.002$^\circ$] and [178.876$^\circ$:179.99$^\circ$] with a 0.124$^\circ$-step. The last two ranges were chosen to explore nearly-coplanar configurations avoiding the higher collision probability inherent to the two dimensional problem. In addition to the three main characteristic drift times, we simulated four additional values ($\tau=2$, 4, 6 and 8 $\times 10^5$) with an inclination step of 5$^\circ$ in order to determine if capture likelihood  evolves smoothly with respect to drift time which it does. 

For each inclination and each eccentricity standard deviation  $\sigma_e=0.01$, $0.1$, $0.3$ and $0.7$, we generate  a statistical ensemble of 1000 particles with random uniform distributions for mean longitudes and nodes, as well as a Rayleigh distribution for eccentricity. The study thus totals $1\,676\,000$ numerical simulations of resonant capture. 
Thirty resonances are monitored: 5:1, 4:1, 3:1, 5:2, 7:3, 2:1, 9:5, 7:4, 5:4, 8:5, 3:2, 7:5, 4:3, 5:4,  6:5, 1:1, 5:6, 4:5, 3:4, 5:7, 2:3, 5:8, 3:5, 4:7, 5:9, 1:2, 3:7, 2:5, 1:3 and 1:4. As in Paper I, they were chosen as the main first, second, third and fourth order (prograde) resonances. We note however that resonance order for retrograde motion is higher at the same nominal location. 

{  One of the important initial parameters in our simulations is the integration timespan. Only the correct choice of this parameter will allow us to meaningfully compare the behaviour of the various resonances with the different drift rates. Setting up a fixed integration time span for all simulations --that covers the longest drift time-- is not adequate. To see this, suppose that the planet suddenly disappeared from the system, particles with different radial drift rates starting at $3 a_p$ will end up in different locations and would not therefore have crossed the same  physical space nor the same resonance web unless the integration time span is long enough so that particles with the slowest drift rate can reach the star's surface. In that case,  all particles if unperturbed would have travelled across the same space (and removed if they collided with the star) but those captured quickly in resonance with the fastest drift rates will artificially increase their eccentricity to unity and end up colliding with the star.  In fact, if the simulation time span is long enough, all captured particles will be removed regardless of the drift time. A very long time span will also remove the resonance-free particles (see next section) and there would be no way to measure their importance with respect to captured and ejected particles. To remedy these problems, we do not choose an arbitrary integration time span but impose physical boundaries on the simulations so that regardless of drift rate, a given particle if unperturbed would travel from $3 a_p$ to an inner location that we choose to be a semimajor axis of $0.246a_p$. This value reflects our interest in the transport of asteroids and centaurs from the outer to the inner the solar system (for $a_p=5$ AU, the inner boundary is at  1.23 AU). With the drag force that we chose this implies an integration time span  2.5 times the drift time under study preventing particles from being lost because of an artificial increase in eccentricity if they were captured or because they precipitated into the star after having resisted resonance capture by the planet (see next section).  }

{
\section{Orbital outcomes}
For a given particle's initial conditions, five orbital outcomes are possible: capture, ejection, collision with the star, collision with the planet and stable resonance-free evolution.

\subsection{Resonance capture}
A particle is considered captured in mean motion resonance if it survives the full simulation duration. The simulation's time span that allows a particle unperturbed by the planet to reach  a semimajor axis of 0.246 was chosen so as to give an opportunity to those particles that were not captured or were released from resonance to be trapped at another location. The capture mechanism in high order resonance for moderate initial eccentricity and inclination was examined in Paper I and shown to present three distinct stages:  capture occurs first in the lowest order inclination resonance as its Hamiltonian structure far from the nominal resonance location is identical to that of a first order resonance (for details see the Hamiltonian model described in the Appendix of Paper I). The second stage sees the activation of the Kozai-Lidov resonance for a duration that depends on the drift time (the longer the drift time the longer the Kozai-Lidov episode) and ushers capture in high order eccentricity resonance that constitutes the third stage. For larger eccentricities, the particles may be or enter the Kozai-Lidov resonance before mean motion resonance capture. We do not produce examples of orbital evolution of captured particles as that was done in Paper I. 

\subsection{Ejection}
Ejection is seen to occur in our simulations when particles captured in mean motion resonance do not survive close encounters with the planet. Such close encounters are favoured by the eccentricity increase during capture. This state is independent of the semi-major axis drift rate. Fig. 1 shows two such examples. For the fastest drift rate, $\tau=10^5$ (studied in Paper I), we have chosen one of the few initial conditions that led to ejection for $I_0=7.5^\circ$ and $\sigma_e=0.01$ (initial eccentricity $e_0=0.00958$). The ejection rate at such an initial inclination is quite low as will be shown in the next section. The particle is trapped in the 1:2 resonance and its eccentricity increases to $\sim 0.5$ where a brief passage through the Kozai-Lidov resonance lowers inclination further. Close encounters with the planet occurs when $e\sim 0.6$ and $I\sim 4^\circ$ and destabilise the orbit after a short time. For the slowest drift rate, $\tau=10^7$, the particle's orbit was taken from the ensemble $I_0=2.5^\circ$ and $\sigma_e=0.01$ (initial eccentricity $e_0=0.00827$).  The ejection rate at such an initial inclination is quite significant as will be shown in the next section. The particle is captured in the 1:3 resonance and unlike the previous example with a first order resonance, capture follows the three stage process where the lowest order inclination is the first to capture the particle as evidenced by the inclination growth. After its release from the Kozai-Lidov resonance and capture in the eccentricity resonance, the particle suffers several close encounters with the planet starting from $e\sim 0.7$ and is rapidly ejected from the system. 

\subsection{Comparative outcome statistics}
 The relative importance of the various orbital outcomes in terms of the ratio of the particle number in a given outcome to the total number of particles  in an ensemble ($=1000$) is shown in Figure 2 for the fastest and slowest drift rates ($\tau=10^5$ and $10^7$). Only the fraction of planet colliding particles is not shown as our simulations indicate that such outcomes are quite rare ($<0.2\%$ for $\tau=10^5$ and $<0.1\%$ for $\tau=10^7$.) 
 
 For the fastest drift rate, the dominant outcomes are found to be capture, followed by ejection, resonance-free evolution and collision with the star. For the slowest drift rate, the latter two positions are exchanged and very few particles remain resonance free.  
 The significant lack of stable captured particles is directly correlated with the increase of ejected particles regardless of the drift rate. As mentioned in the previous section, ejected particles were captured in mean motion resonance, increased their eccentricity during the capture episode and suffered close encounters with the planet. Resonances are most susceptible to close planet encounters when motion is prograde. The stability against planet encounters of  retrograde resonances originates in the physical setting at the encounter specifically its duration and relative velocity. Particles on retrograde orbits encounter the planet with higher relative velocity during a shorter time than  prograde orbits do.  For the fastest drift rate, resonance-free particles are more present for prograde small eccentricity orbits and retrograde large eccentricity orbits. For the slowest drift rate, a similar trend is observed but the overall fraction of resonance-free particles is negligible.  
 
Star collisions peak for polar and nearly polar retrograde orbits. This occurs because once captured in resonance such orbits are unlikely to encounter the planet and increase their eccentricity to near unity where they collide with the star.  This brings us back to the significance of what is measured as a capture fraction. In our simulations, only particles that survive the full duration of the simulation and end up in a resonant state are termed captured in resonance. Particles that were captured in resonance and subsequently ejected by close encounters with the planet, or suffered a collision with the star are not counted as captured. Such a definition of capture as  stable long-lived capture allows us to assess the robustness of the resonances that we consider with a possible future application to the dynamical origin and stability of the Centaurs and Damocloids in high inclination orbits in the solar system. In the next section, we specialise in the long-lived capture orbital states and analyse the effect of the radial drift rates on the efficiency of mean motion resonance capture as a function of the resonance order and initial inclination.

\section{Resonance capture statistics}
Our choice} of initial conditions and in particular that of the semimajor axis outside the 1:5 resonance with the planet implies that  the measured capture likelihood in a given resonance only indicates conditional capture probability and is in no way absolute (with the exception of the 1:5 resonance) as particles are forced to drift radially and cross the various resonances that may or may not capture them. We also note that since our ensembles for each inclination contains 1000 particles, global {long-lived} capture likelihood is known at best down to 1 part in 1000. The results of our simulations are expanded on in the following sections. 

\subsection{Global long-lived capture statistics} 
The fraction of particles captured in each ensemble expressed in percent is shown in the top row of Fig. 3 as a function of the initial inclination for the three drift rates, $\tau=10^5, \ 10^6$ and $10^7$. Global capture fraction is reduced overall as the drift time is increased. 

For small eccentricity, $\sigma_{\rm e}=0.01$, and beyond $125^\circ$ initial inclination, capture is nearly independent of drift rate. The capture peak at zero inclination for nearly circular orbits disappears in favour of a new one at 5$^\circ$ as drift time increases. A deep gap appears between $15^\circ$ and $45^\circ$ as well as another around $110^\circ$ that deepen with drift time. Beyond $125^\circ$ initial inclination, capture fraction is almost unaffected by the drift rate except for a small gap near $175^\circ$ for the slowest drift rate. As eccentricity is increased, the gaps' edges become less sharp and capture fraction is reduced further beyond $125^\circ$.  Capture of nearly coplanar retrograde orbits for the drift rate $\tau=10^7$ is reduced to a quarter of its value for the base drift time $\tau=10^5$ of Paper I. 
{The statistics of high order resonance capture examined below explains why the global capture fraction decreases as the drift time becomes longer.  These resonances are more efficient at capture for slow drift times but they are also more susceptible to close planet encounters than first order resonances.  Disruptive close planet encounters not only make particles leave the resonance but they eject them from the system much like the examples shown in Fig. 1. As particles on retrograde orbits encounter the planet with higher relative velocity during a shorter time than  prograde orbits do, high order retrograde resonances are less affected by close planet encounters. Therefore,} regardless of eccentricity and drift rate, we confirm the finding of Paper I that retrograde resonance capture are more efficient for retrograde orbits than prograde orbits. 

\subsection{Coorbital resonance}
The fraction of particles captured in the coorbital 1:1 resonance among those captured at all is shown in the bottom row of Fig. 3 as a function of the initial inclination. Being the strongest mean motion resonance in the system, it is interesting that we find it marginally affected by the longer drift rates except for some specific domains. For nearly circular orbits, the smaller drift rates may double the capture likelihood of the small peak at $80^\circ$ (compare the curves of $\tau=10^5$ and $10^6$), lower that of the peak at $110^\circ$ and reduce by a few degrees the initial inclination extent of unit capture probability for nearly  coplanar retrograde orbits. For moderate eccentricities ($\sigma_{\rm e}=0.10,\, 0.30$), changes are marginal except for $180^\circ$ initial inclination and  $\sigma_{\rm e}=0.30$ where capture likelihood jumps from $\sim 40\%$ with $\tau=10^5$ to $\sim70\%$ with $\tau=10^7$. For highly eccentric orbits ($\sigma_{\rm e}=0.70$), capture is enhanced significantly by the longest drift rate for initial inclinations around $110^\circ$ and $180^\circ$. At the former, there appears a strong probability peak about $15^\circ$-wide that follows a zero capture probability range. At $180^\circ$ initial inclination, capture likelihood increases from $\sim25\%$ with $\tau=10^5$ to $\sim40\%$ with $\tau=10^7$. As with global capture, retrograde orbits are far more efficiently trapped by the coorbital resonance than prograde orbits. A detailed study of the workings of the coorbital resonance may be  found in our recent work \cite{MoraisNamouni16}.

\subsection{First order outer resonances}
The fraction of particles captured in  first order outer resonances among those captured in resonance by the planet is shown in  Fig. 4 as a function of the initial inclination. The resonance rows are ordered vertically according to their encounter with the particles. Only those reaching or exceeding $\sim20$\% capture fraction are shown. They number three and reach a maximum of $\sim 100\%$ for 1:2 at $\sigma_{\rm e}=0.01$,  $\sim 30\%$ for 2:3 at $\sigma_{\rm e}=0.10$, and $\sim 30\%$ for 5:6 at $\sigma_{\rm e}=0.01, 0.10$ and $0.30$. 

For nearly circular orbits ($\sigma_{\rm e}=0.01$), the 1:2 capture likelihood is reduced for nearly coplanar prograde orbits as the drift time becomes longer and drops from $\sim100\%$ for $\tau=10^5$ to $\sim 20\%$  for $\tau=10^7$. Smaller peaks and gaps punctuate the capture fraction curve especially one at an initial inclination of $45^\circ$  where likelihood peaks at $\sim65\%$. The efficiency of the 1:2 resonance at capturing retrograde orbits (initial inclination $>120^\circ$) is seriously reduced for longer drift times by the outer higher order resonances that trap particles first and prevent them traveling further towards the planet. For large eccentricity ensembles ($\sigma_{\rm e}=0.30,\ 0.70$), longer drift times tend to enhance capture likelihood threefold in the vicinity $120^\circ$ initial inclination.

For moderate  to large eccentricity ($\sigma_{\rm e}=0.10,\ 0.30$ and 0.70), capture in the 2:3 resonance is  globally reduced by slower drift rates. For nearly circular orbits ($\sigma_{\rm e}=0.01$) and longer drift times, particles are not trapped in resonance except those with nearly polar orbits and to a lesser extent those with inclinations of $60^\circ$, $110^\circ$ and  $150^\circ$. 
This preference for polar orbits with longer drift rates is displayed by the 5:6 resonances as capture likelihood is increased threefold to fivefold.

\subsection{Second and third order outer resonances}
The fraction of particles captured in  second and third order outer resonances among those captured in resonance by the planet is shown in  Fig. 5 as a function of the initial inclination. The resonance rows are ordered vertically according to their encounter with the particles. Only 1:3, 5:7, 1:4 and 2:5  reach or exceed  $\sim20$\% capture fractions. The effect of longer drift times on second order planar resonances studied by \cite{Quillen06} may be seen on the initially nearly coplanar orbits that are captured in 1:3 resonance. For the initially nearly circular orbits' ensembles ($\sigma_{\rm e}=0.01$), only the longest drift rate $\tau=10^7$ increases the capture fraction to nearly 80\% for nearly coplanar orbits. A secondary capture peak rises near $25^\circ$. For larger eccentricity ensembles  ($\sigma_{\rm e}=0.10,\, 0.30$ and 0.70), capture is enhanced only for retrograde orbits with an initial inclination larger than $120^\circ$. The 5:7 second order resonance has an efficiency peak at an initial inclination of $\sim 35^\circ$ for nearly circular orbits but otherwise shows a preference for polar orbits at larger eccentricities. The third order outer resonances 1:4 and 2:5 show some of the most important changes as the drift rate is increased.  {Three narrow capture peaks for nearly circular orbits ($\sigma_{\rm e}=0.01$) at 60\% and 100\% for initial inclinations between $15^\circ$ and $35^\circ$ appear for 1:4 along a much broader one for the initial inclination range [$90^\circ,160^\circ$] rising  to 100\%.} As the initial eccentricity standard deviation  is increased, a single efficiency peak remains for prograde orbits around $40^\circ$ up to $\sigma_e=0.30$ whereas capture of polar and retrograde orbits is enhanced significantly for the former and moderately for the latter. The longer drift time capture fractions of the 2:5 resonances peak at least at 70\% for prograde orbits and $\sigma_{\rm e}\leq 0.30$. For nearly circular orbits an efficiency peak at nearly 80\% appears around an initial inclination of $15^\circ$ but not for initially coplanar orbits. 

{With the longest radial drift rate $\tau=10^7$ capture of prograde orbits occurs mainly  in high order resonances.  Although the results of this section refer to particles that survived the full integration timespan, it is worth noting that the large fraction of ejected particles on prograde orbits with the longest drift time were captured by high order resonances.  The susceptibility of such resonances to  close planet encounters accounts for the capture asymmetry of prograde and retrograde resonances as the latter are less disrupted because encounters occur for a shorter duration and at a larger velocity with respect to prograde motion. }

\subsection{Fourth order outer resonances}
The fraction of particles captured in  second and third order outer resonances among those captured in resonance by the planet is shown in  Fig. 6 as a function of the initial inclination.  Only 1:5 and 3:7  reach or exceed  $\sim20$\% capture fractions. The case of the 1:5 resonance is interesting as it is the only resonance for which an absolute capture fraction may be known since it is the first major resonance to be encountered by all particles. 
{Capture for nearly coplanar circular orbits is a rare event even with the slowest drift rate.} For the shortest drift time $\tau=10^5$, this resonance captured prograde orbits more efficiently if their initial eccentricities were very small ($\sigma_{\rm e}=0.01$) or very large ($\sigma_{\rm e}=0.70$) and  retrograde orbits only at large eccentricity ($\sigma_{\rm e}=0.70$). For all but the largest initial eccentricities, this behaviour is modified by the larger drift rates as polar orbits are singled out for efficient capture, retrograde capture is enhanced but prograde orbits not captured at all. For the largest eccentricities ($\sigma_{\rm e}=0.70$) capture is enhanced overall with a peak at 45\% for prograde orbits with an initial inclination  of $30^\circ$ and a wider one around $90^\circ$ at 80\%. This reinforces our earlier conclusion in Paper I that resonance order at arbitrary inclination (and eccentricity) is not  necessarily a good indicator for capture efficiency. 
The fourth order 3:7 outer resonance shows an equally interesting behaviour: apart from the much enhanced capture of polar orbits as the drift time is increased (fivefold for $\sigma_{\rm e}=0.10$ and $\tau=10^6$ and tenfold for $\tau=10^7$), singular narrow efficiency  peaks show up for prograde orbits at locations unrelated to that of the faster drift rate $\tau=10^5$ at $45^\circ$. Moreover, capture in retrograde resonances is reduced and rendered impossible at certain inclinations (e.g. [$100^\circ,120^\circ$] for $\sigma_{\rm e}=0.70$ and $\tau=10^7$). 

\subsection{First order inner resonances}
As capture in high order outer resonances is found to be enhanced by the longer drift times, one expects that the first order (and strongest) inner resonances will capture less particles. The reality is a little different (Fig. 7). For all but the high eccentricity ensembles, capture is mainly suppressed at most inclinations except in a $15^\circ$-wide range adjacent near $75^\circ$ for the 6:5, 5:4 and 3:2 resonances and $65^\circ$ for the 2:1 resonance (with $\tau=10^7$) where capture is significantly enhanced. For the largest eccentricities ($\sigma_{\rm e}=0.70$), in addition to increasing the width of the previous domain from $15^\circ$ to $30^\circ$, several capture peaks are found for smaller initial inclinations.  

\subsection{High order inner resonances}
The effect of longer drift times on resonance capture is perhaps most surprising for high order inner resonances as particles are set out far outside the planet's orbit. The resonance capture fractions are shown in Fig. 8 and 9. For the drift rate of Paper I, $\tau=10^5$, only the inner  third order 5:2 resonance showed a large capture likelihood $\sim 35\%$ for initially nearly circular orbits centred around the initial inclination of $55^\circ$. With the longer drift rates of $\tau=10^6$ and $10^7$, six resonances show a capture fraction in excess of $20\%$. They are second order 7:5, 5:3 and 3:1, fourth order 9:5 and 7:3 and third order 5:2. The longer drift rates generally tend to increase the amplitude of the capture peak of maximum efficiency  and narrow its width. For all but the largest eccentricities, additional capture peaks are present for smaller prograde inclinations.   The 7:5 resonance doubles its capture fraction peaking at $75^\circ$ initial inclination. The 5:3 capture fraction around the same location is increased  threefold to fivefold depending on the eccentricity standard deviation and drift time. The 3:1 resonance develops very narrow capture peaks for prograde orbits at low to moderate eccentricities and additional peaks for retrograde large eccentricity orbits. Fourth order resonances are similar to second order resonances except for the precise location of their peak efficiency and their capture amplitudes that tend to be larger at moderate to high eccentricities. The third order 5:2 resonance shows the  largest amplitudes among inner resonances exceeding 90\% for $\tau=10^6$ and $\sigma_{\rm e}=0.01$ and 80\% for $\tau=10^7$ and $\sigma_{\rm e}=0.30$ peaking around $50^\circ$ to $55^\circ$ initial inclination.  Although the global fraction of captured particles at these initial inclinations is small (a few to 10\%), most of it will end up inside the planet's orbit at the 5:2 resonance. More generally, we may conclude that in the initial inclination range [$50^\circ$,$80^\circ$], there seems to be corridors for particles to drift past the planet's orbit and be captured in high order inner resonances. Examination of the final states in that initial inclination range reveals that particles with moderate eccentricities and an initial inclination lower  than $55^\circ$ tend to remain on prograde orbits.  Otherwise orbits may end up significantly inclined and tend to cluster around the Kozai-Lidov resonance with $\omega=0^\circ$ and $180^\circ$.

\section{Conclusion}
Our extended numerical simulations of resonance capture show that the effect of the radial drift rate is mostly seen on the specific capture ability of individual resonances. {In agreement with Paper I}, retrograde orbits are overall much more likely to be captured in resonance than prograde orbits. {The physical explanation is related to the disruption caused by planetary encounters that favours retrograde orbits as particles encounter the planet for a shorter duration with a higher relative velocity with respect to prograde orbits. The retrograde versus prograde resonance  capture asymmetry } has recently been confirmed in simulations about the origin of Halley-type comets and its possible relation to the evolution of long period comets \citep{Fernandez16}.

Longer radial drift times favour capture in high order resonances especially for low eccentricity orbits at low initial inclinations and moderate to high eccentricity for polar and sub-polar initial inclinations confirming the findings of \cite{Quillen06} as far as two dimensional capture is concerned.  {As 
high order resonances are more affected by disruptive planetary encounters because of their intrinsic dynamical weakness, longer radial drift times reduce the overall capture tally especially for prograde orbits. In this regard, our results also imply that the drift rates at which higher order retrograde resonances are more efficient at resonance capture may still be larger than the longest drift time employed in this study. 
In effect, the capture likelihood of nearly circular orbits with initial inclinations from $150^\circ$ to $180^\circ$ is unaffected by changing the drift times (we used) and their capture occurs always in the 1:2 resonance instead for instance of the 1:3 or 1:4 as for prograde motion.

We have uncovered stable dynamical corridors  that guarantee safe passage to particles starting from outside the 1:5 resonance of a given planet all the way to its inner 5:2 resonance where they are finally captured. As our study is based on the three-body problem, we are not able to predict the behaviour of such corridors for a multi-planet system and in particular the solar system with its four outer planets. More simulations are required to ascertain whether such corridors exist and how those pertaining to different planets combine. 

Our main motivation for this paper  and its predecessor, is the origin of Centaurs and Damocloids on retrograde orbits. These objects} are known to follow two types of dynamical evolution \citep{BaileyMalhotra09,VolkMalhotra13}: random walk and resonance sticking. In the first type, orbits have the shorter lifetimes of order $10^7$ years as they are dominated by frequent close encounters with the giant planets whereas in the second type they are long lived and may survive $10^8$ years.  For the latter type, radial drift is mainly caused by weak chaos originating from the overlapping of resonances of the four giant planets in the solar system. The insight we gained from exploring resonant capture at arbitrary inclination for migration timescales comparable to the chaos generated drift rates will help us elucidate the significance of the bipolar nature of the Centaurs' dynamical evolution and its possible link to their original reservoirs in the solar system.

\section*{Acknowledgments}
The numerical simulations in this work were performed at the Centre for Intensive Computing  `M\'esocentre {\sc sigamm}'  hosted by the Observatoire de la C\^ote dÕAzur.

\begin{figure*}
\begin{center}
\includegraphics[width=85mm]{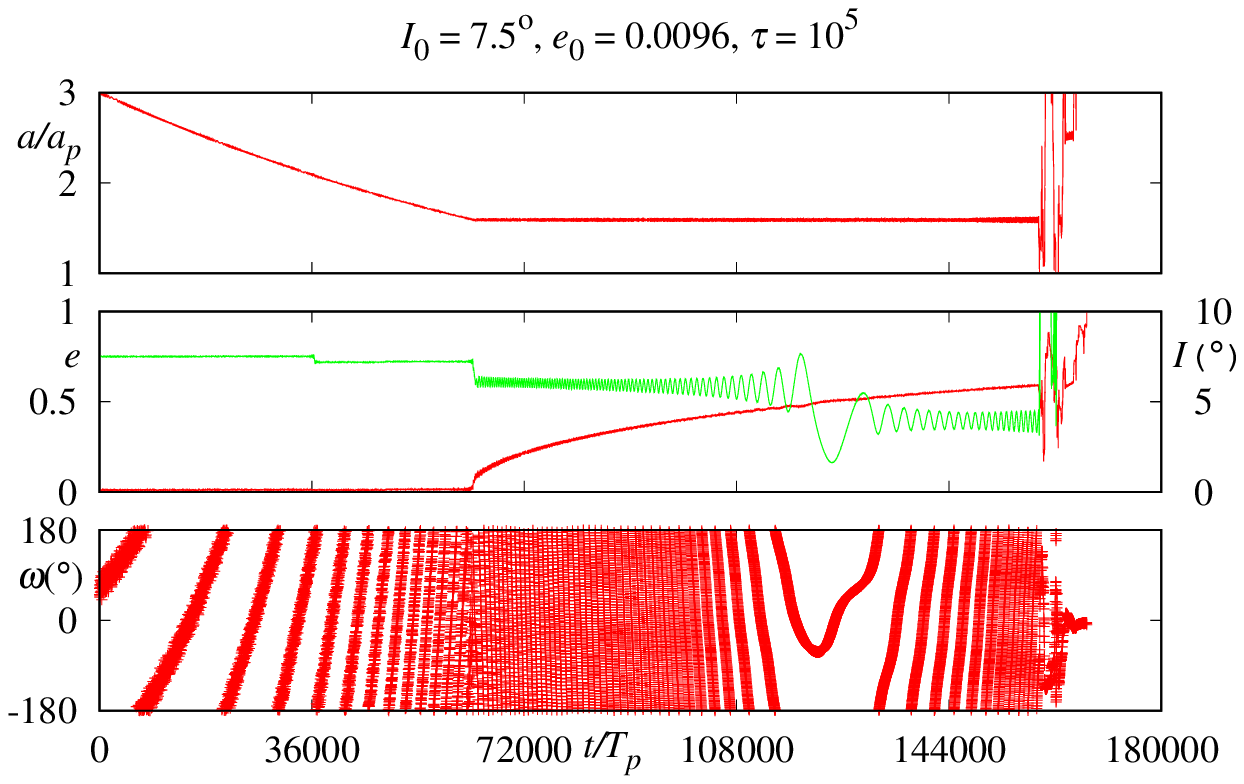}
\includegraphics[width=85mm]{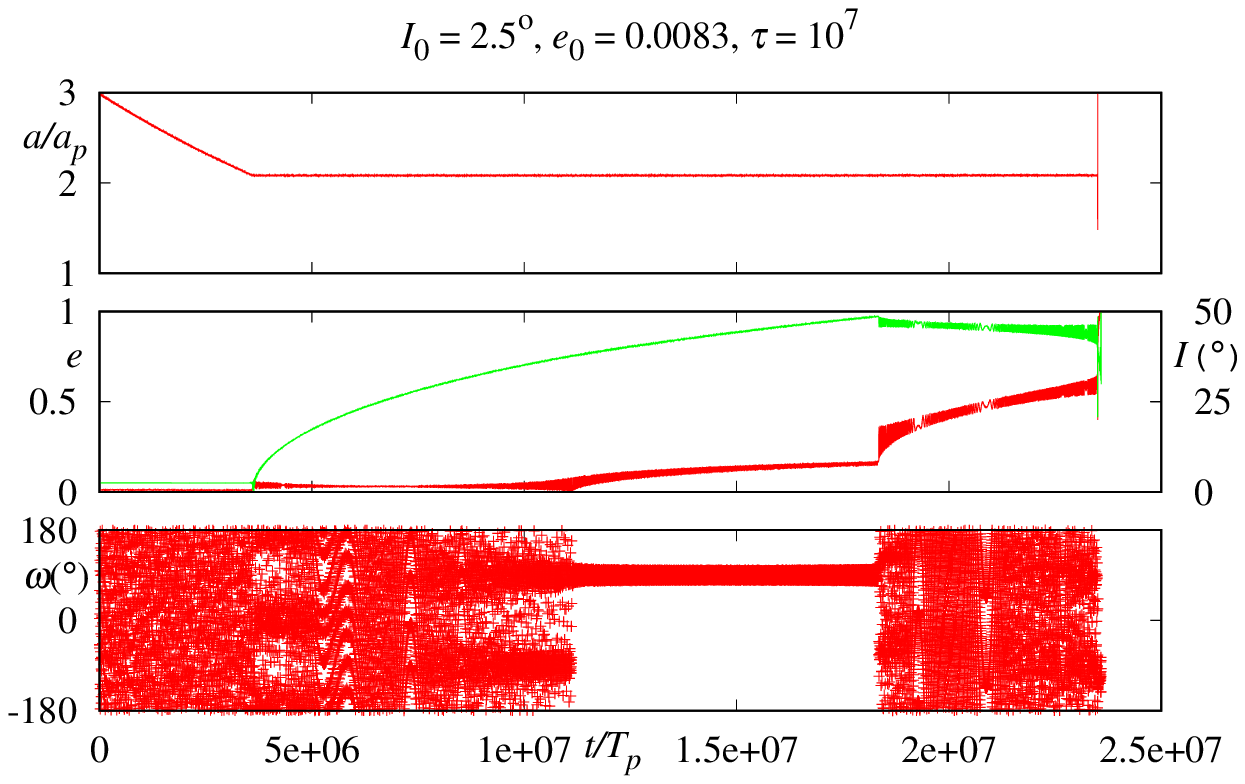}
\caption{Typical time evolution of ejected orbits for the fastest (left panels) and slowest (right panels) drift rates. Time $t$ is normalized to the planet's period $T_p$.   The semi-major axes' ratio of the particle's and the planet's $a/a_p$  is shown in the top panels, eccentricity $e$ (solid red, left vertical axis) and Inclination (dashed green,  right vertical axis) in the middle panels and the argument of pericentre (solid red) in the bottom panels.}
\end{center}
\end{figure*}

\begin{figure*}
\begin{center}
\includegraphics[width=62mm]{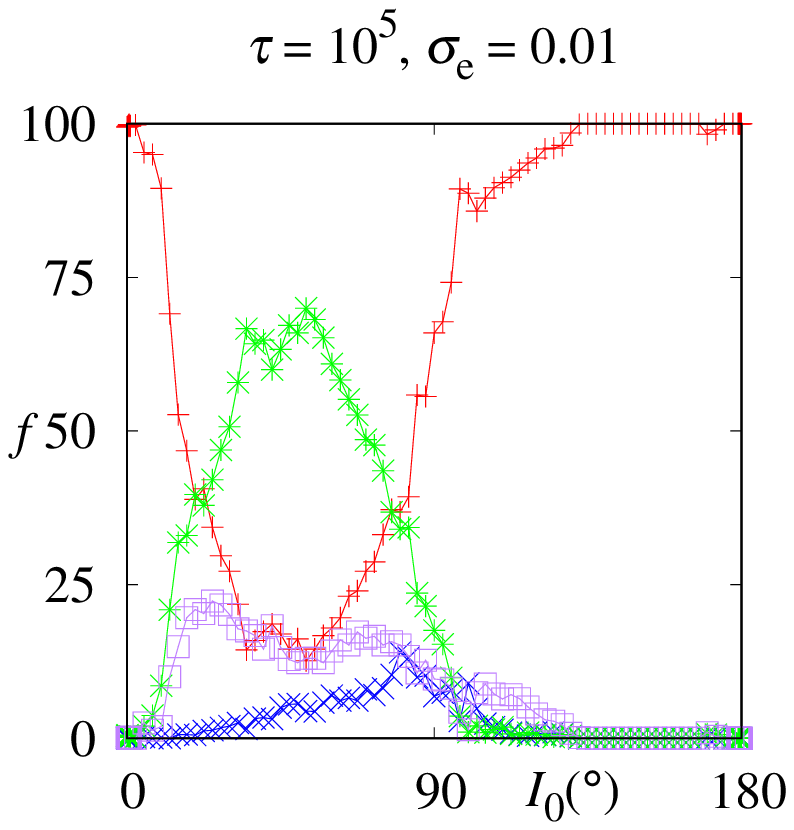}\hspace{-23mm}\hspace{-2mm}
\includegraphics[width=62mm]{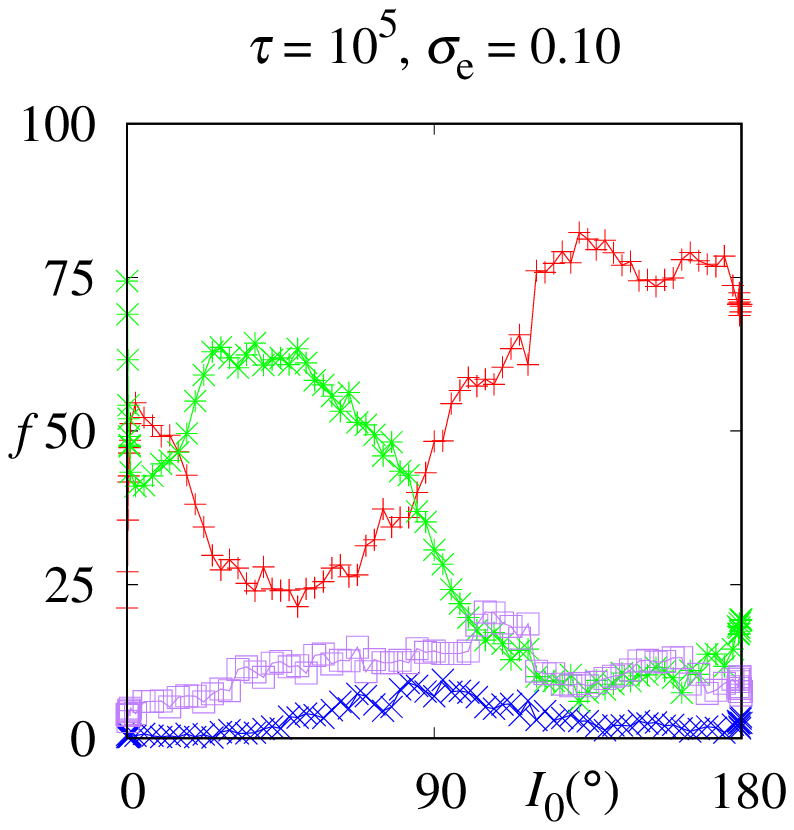}\hspace{-23mm}\hspace{-2mm}
\includegraphics[width=62mm]{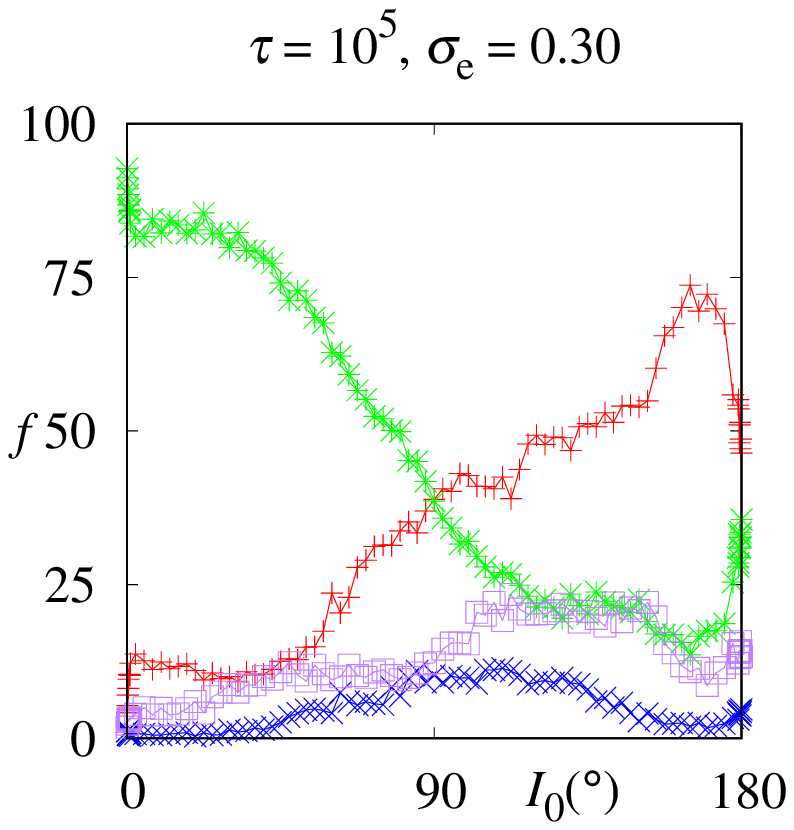}\hspace{-23mm}\hspace{-2mm}
\includegraphics[width=62mm]{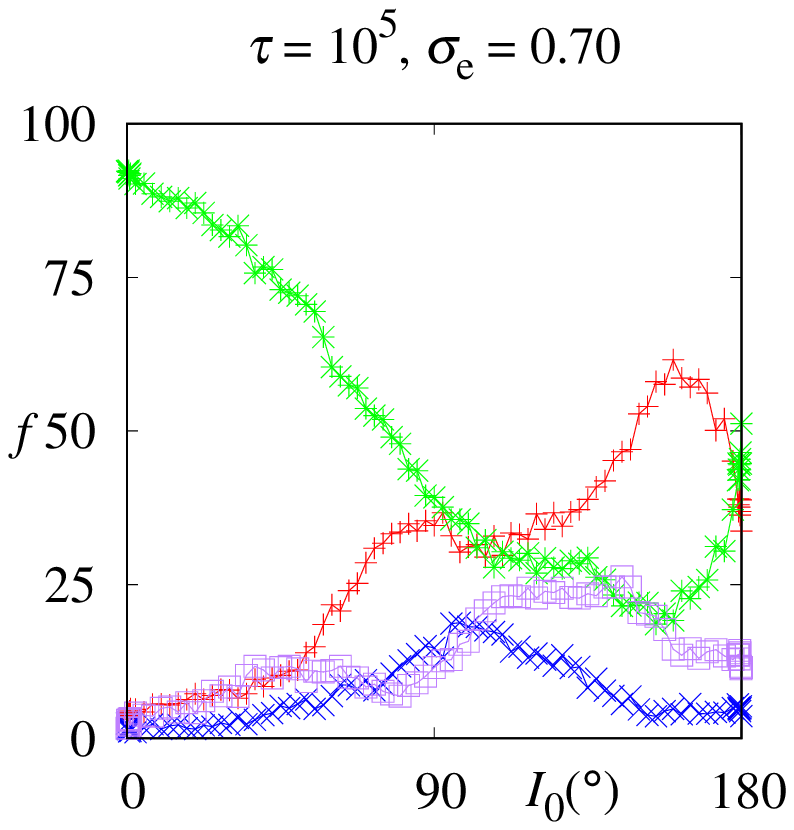}\hspace{-23mm}\\
\includegraphics[width=62mm]{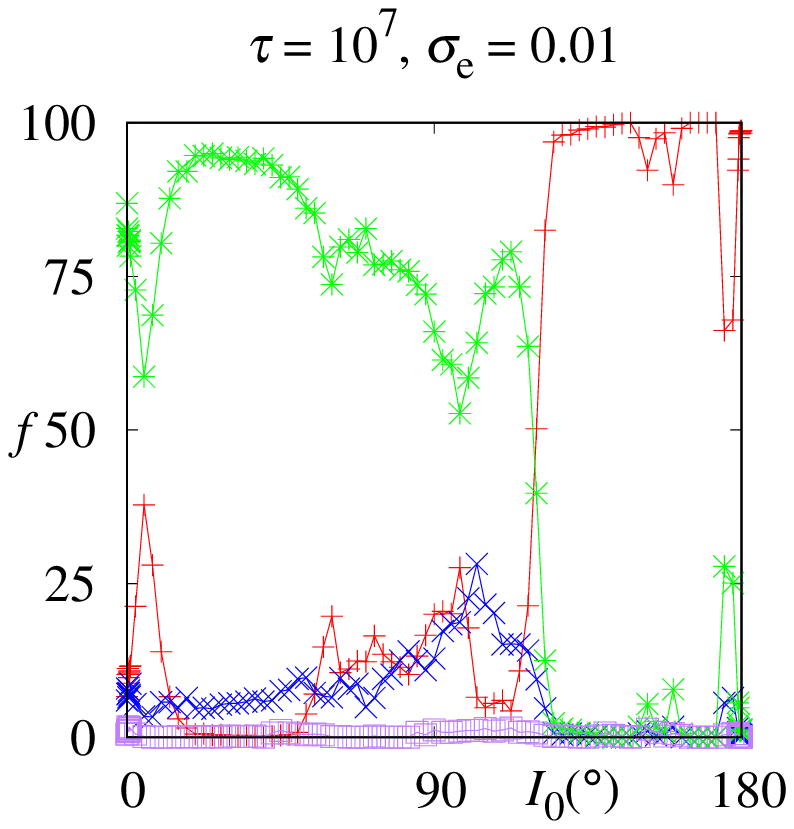}\hspace{-23mm}\hspace{-2mm}
\includegraphics[width=62mm]{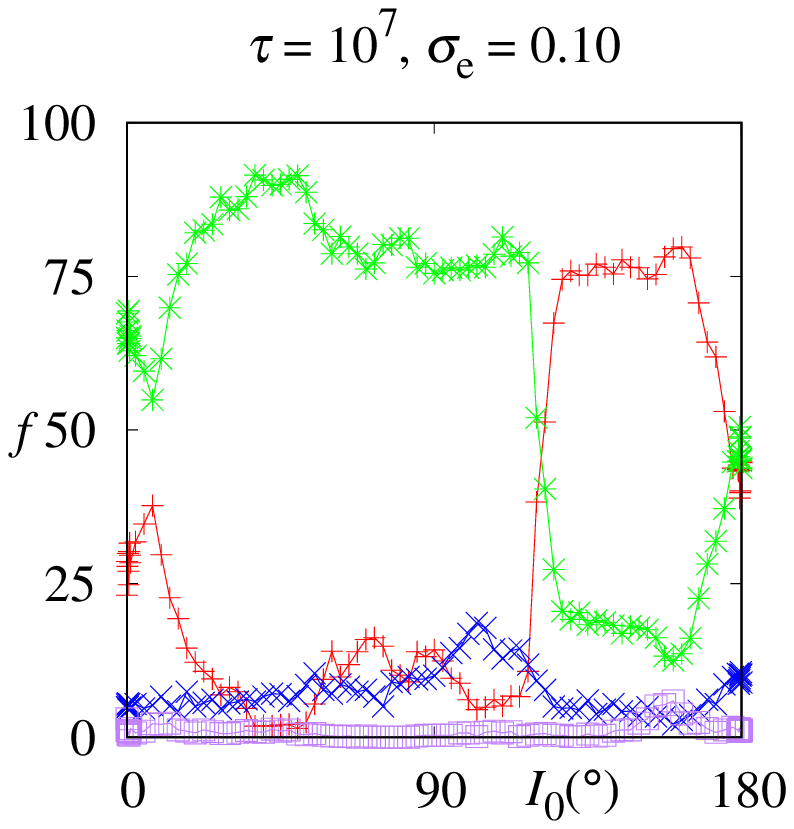}\hspace{-23mm}\hspace{-2mm}
\includegraphics[width=62mm]{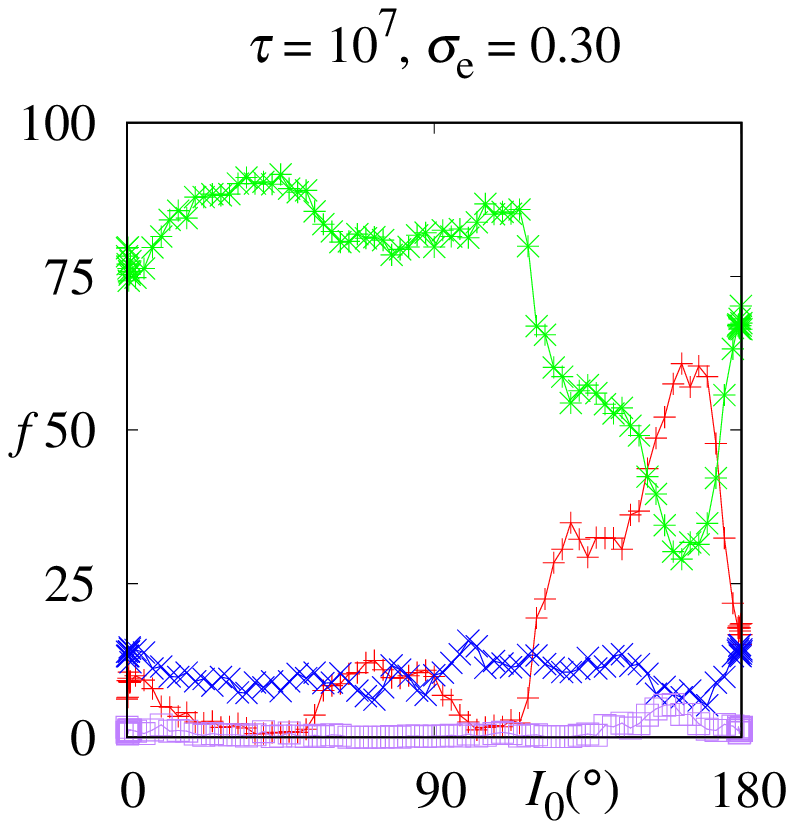}\hspace{-23mm}\hspace{-2mm}
\includegraphics[width=62mm]{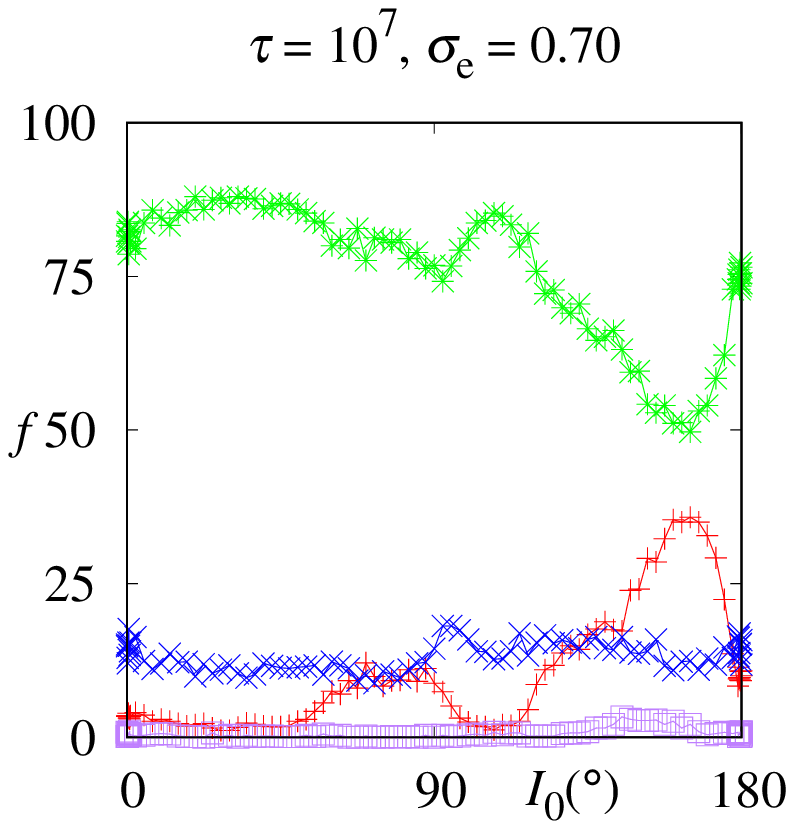}\hspace{-23mm}\\
\caption{Main four outcome particle fractions $f$ (in $\%$) as a function of initial inclination for the fastest (top panels) and slowest (bottom panels) drift rates and each eccentricity standard deviation $\sigma_e$.  The curves correspond to capture (red $+$), ejection (green $*$), solar collision (blue $\times$) and stable resonance-free particles (purple $\square$). }
\end{center}
\end{figure*}

\begin{figure*}
\begin{center}
\includegraphics[width=62mm]{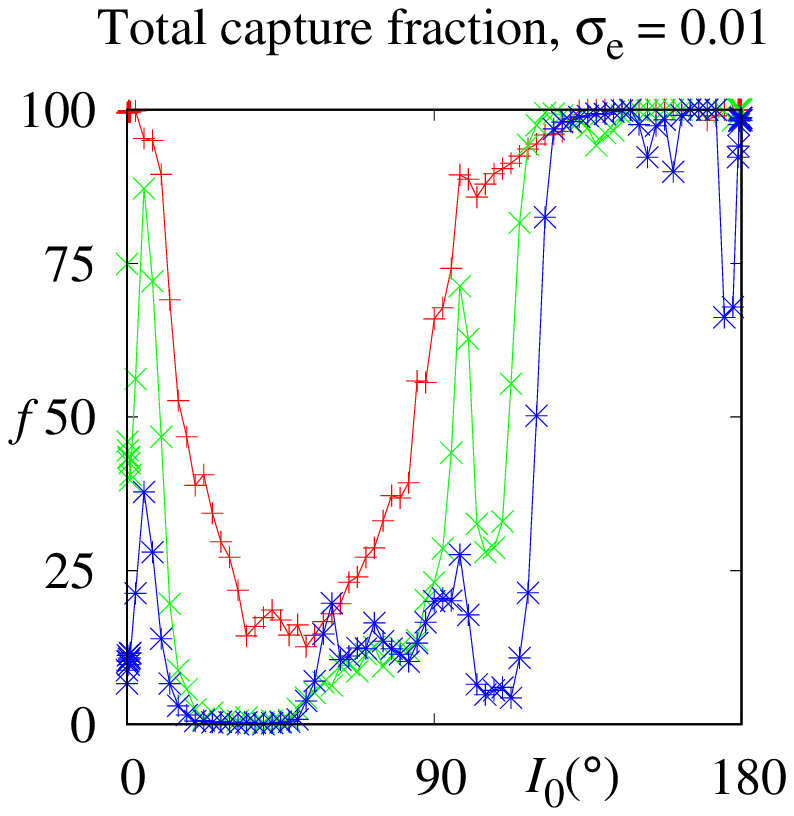}\hspace{-23mm}\hspace{-2mm}
\includegraphics[width=62mm]{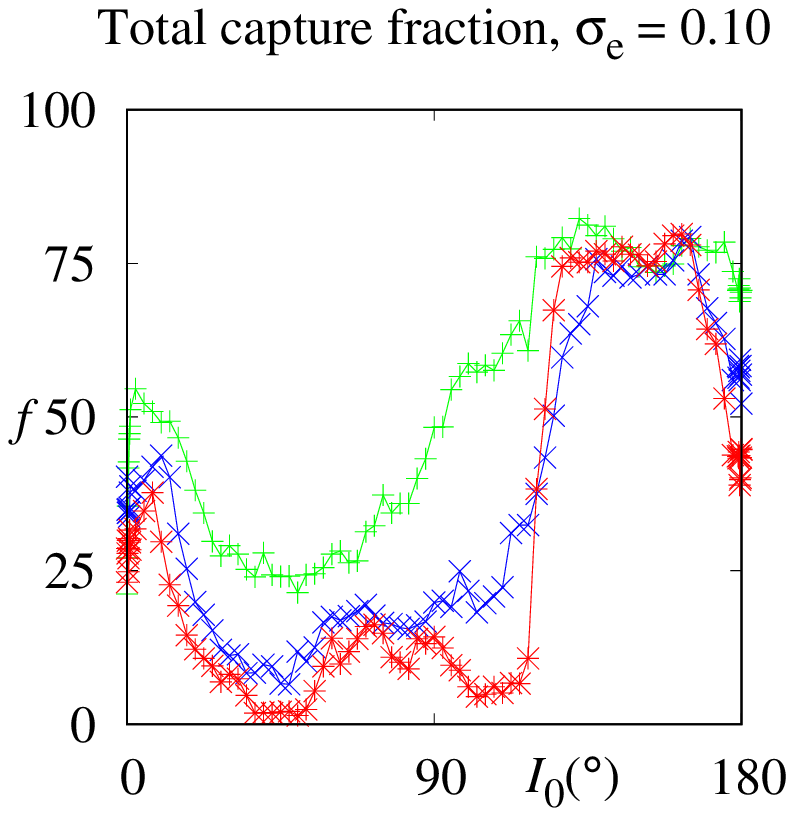}\hspace{-23mm}\hspace{-2mm}
\includegraphics[width=62mm]{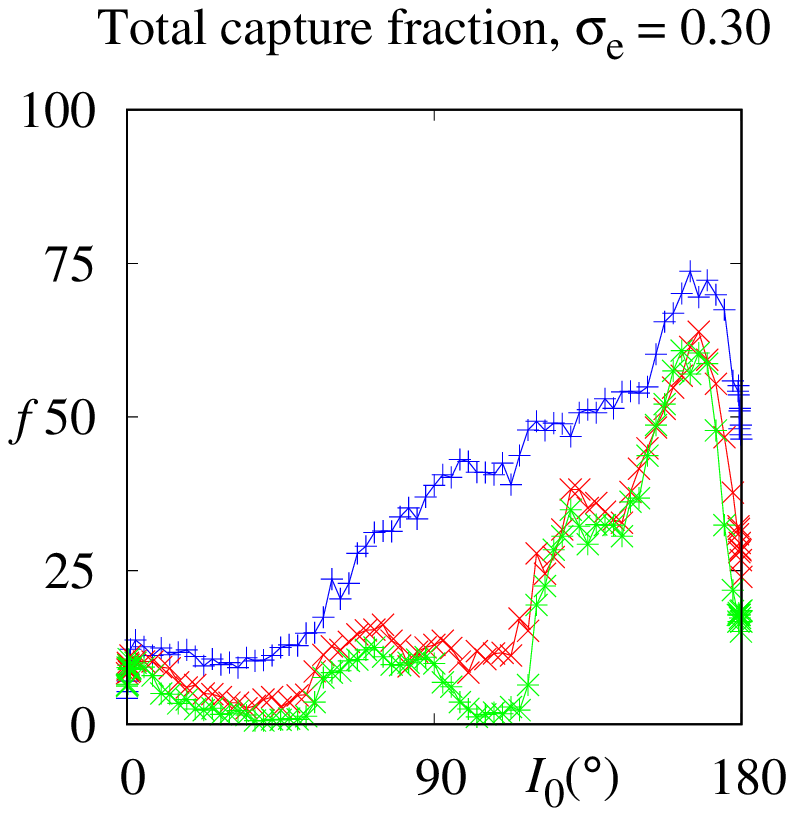}\hspace{-23mm}\hspace{-2mm}
\includegraphics[width=62mm]{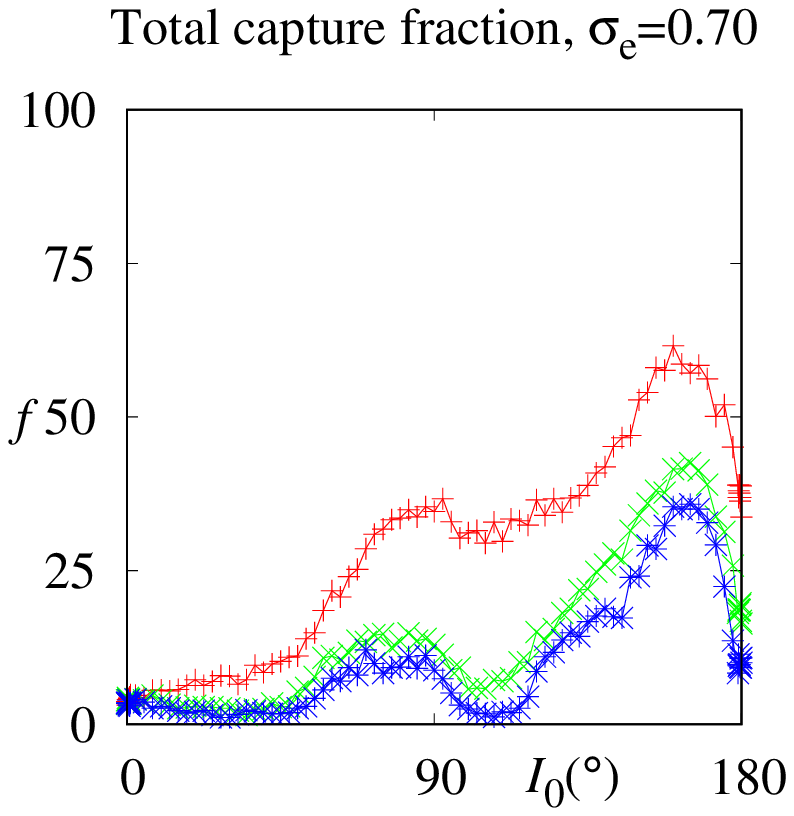}\hspace{-23mm}\\
\foreach \x in {1t1}
{ 
\includegraphics[width=62mm]{fr01-\x .eps}\hspace{-23mm}\hspace{-2mm}
\includegraphics[width=62mm]{fr10-\x .eps}\hspace{-23mm}\hspace{-2mm}
\includegraphics[width=62mm]{fr30-\x .eps}\hspace{-23mm}\hspace{-2mm}
\includegraphics[width=62mm]{fr70-\x .eps}\hspace{-23mm}\\
}
\caption{Total capture  (upper row) and coorbital capture  (lower row) fractions  $f$ as a function of initial inclination for the four eccentricity standard deviations $\sigma_e=0.01, \ 0.10,\ 0.30,\ 0.70.$ In each panel, three curves are plotted  corresponding to the drift rates: $10^5$ yr (solid red $+$), $10^6$ (dashed green $\times$), $10^7$ (dotted blue $\star$). }
\end{center}
\end{figure*}

\begin{figure*}
\begin{center}
\foreach \x in {1t2,2t3,5t6}
{ 
\includegraphics[width=62mm]{fr01-\x .eps}\hspace{-23mm}\hspace{-2mm}
\includegraphics[width=62mm]{fr10-\x .eps}\hspace{-23mm}\hspace{-2mm}
\includegraphics[width=62mm]{fr30-\x .eps}\hspace{-23mm}\hspace{-2mm}
\includegraphics[width=62mm]{fr70-\x .eps}\hspace{-23mm}\\
}
\caption{First-order outer resonance capture fractions. Color and symbol codes are the same as in Figure 1.}
\end{center}
\end{figure*}

\begin{figure*}
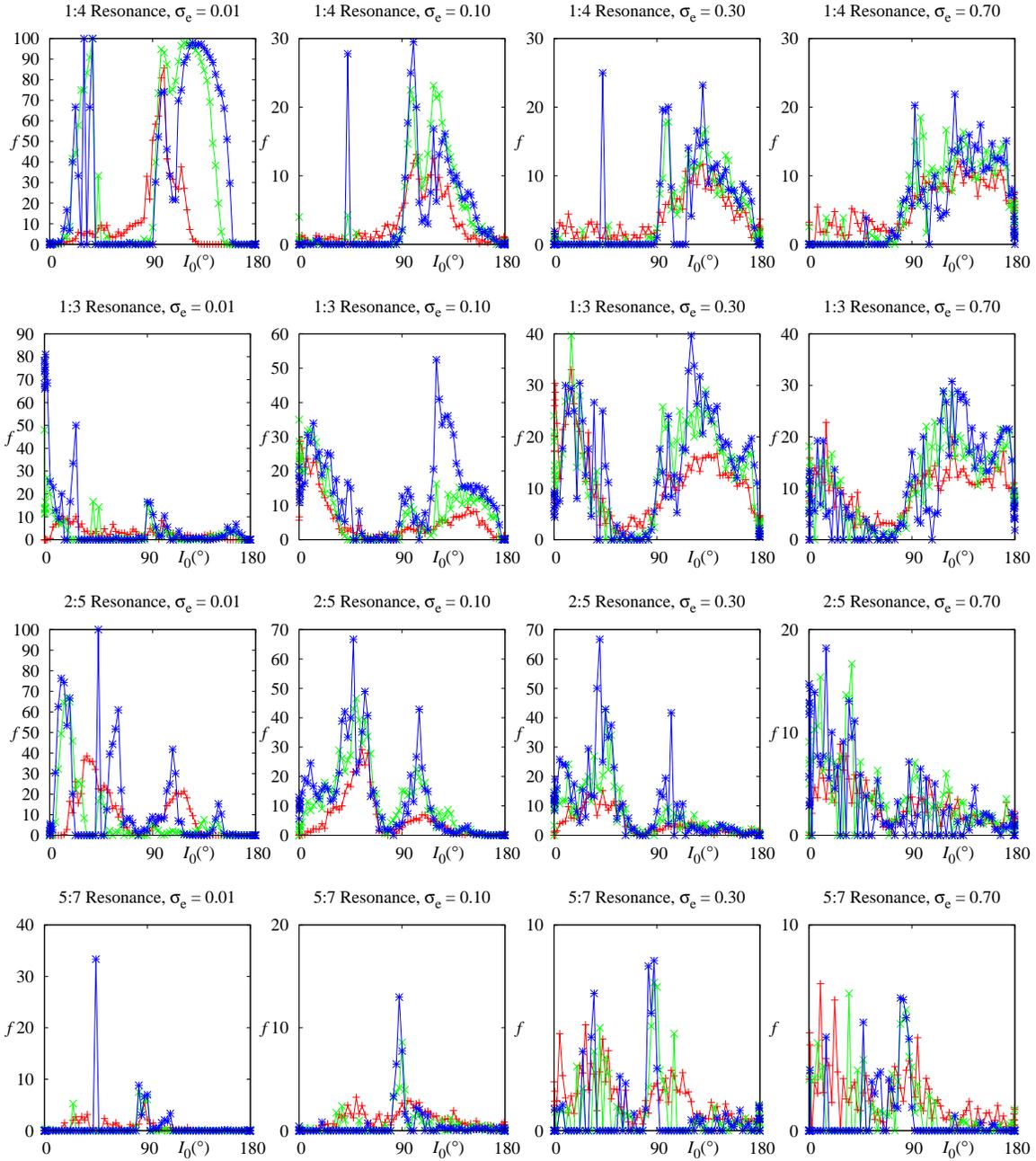

\begin{center}
\foreach \x in {1t4,1t3,2t5,5t7}
{ 
\includegraphics[width=62mm]{fr01-\x .eps}\hspace{-23mm}\hspace{-2mm}
\includegraphics[width=62mm]{fr10-\x .eps}\hspace{-23mm}\hspace{-2mm}
\includegraphics[width=62mm]{fr30-\x .eps}\hspace{-23mm}\hspace{-2mm}
\includegraphics[width=62mm]{fr70-\x .eps}\hspace{-23mm}\\
}
\caption{Second and third order  outer resonance capture fractions. Color and symbol codes are the same as in Figure 1.}
\end{center}
\end{figure*}

\begin{figure*}
\begin{center}
\foreach \x in {1t5,3t7}
{ 
\includegraphics[width=62mm]{fr01-\x .eps}\hspace{-23mm}\hspace{-2mm}
\includegraphics[width=62mm]{fr10-\x .eps}\hspace{-23mm}\hspace{-2mm}
\includegraphics[width=62mm]{fr30-\x .eps}\hspace{-23mm}\hspace{-2mm}
\includegraphics[width=62mm]{fr70-\x .eps}\hspace{-23mm}\\
}
\caption{Fourth order  outer resonance capture fractions. Color and symbol codes are the same as in Figure 1.}
\end{center}
\end{figure*}

\begin{figure*}
\begin{center}
\foreach \x in {6t5,5t4,3t2,2t1}
{ 
\includegraphics[width=62mm]{fr01-\x .eps}\hspace{-23mm}\hspace{-2mm}
\includegraphics[width=62mm]{fr10-\x .eps}\hspace{-23mm}\hspace{-2mm}
\includegraphics[width=62mm]{fr30-\x .eps}\hspace{-23mm}\hspace{-2mm}
\includegraphics[width=62mm]{fr70-\x .eps}\hspace{-23mm}\\
}
\caption{First order inner resonance capture fractions. Color and symbol codes are the same as in Figure 1.}
\end{center}
\end{figure*}

\begin{figure*}
\begin{center}
\foreach \x in {7t5,5t3,9t5,7t3,3t1}
{ 
\includegraphics[width=62mm]{fr01-\x .eps}\hspace{-23mm}\hspace{-2mm}
\includegraphics[width=62mm]{fr10-\x .eps}\hspace{-23mm}\hspace{-2mm}
\includegraphics[width=62mm]{fr30-\x .eps}\hspace{-23mm}\hspace{-2mm}
\includegraphics[width=62mm]{fr70-\x .eps}\hspace{-23mm}\\
}
\caption{Higher order inner resonance capture fractions. Color and symbol codes are the same as in Figure 1.}
\end{center}
\end{figure*}

\begin{figure*}
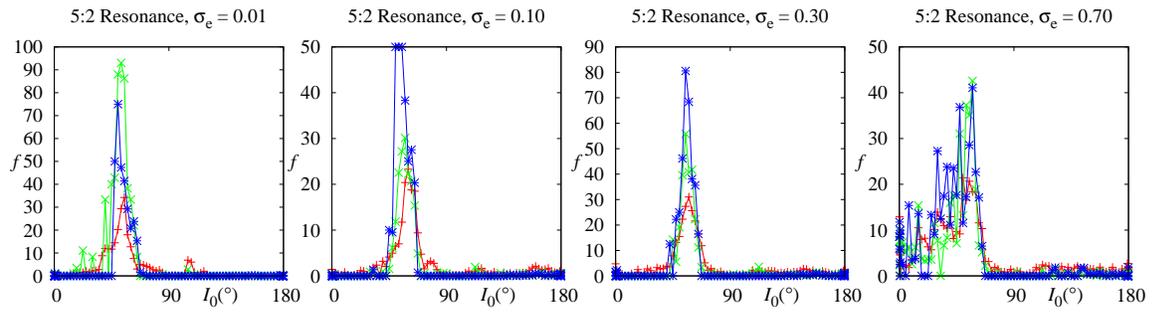

\begin{center}
\foreach \x in {5t2}
{ 
\includegraphics[width=62mm]{fr01-\x .eps}\hspace{-23mm}\hspace{-2mm}
\includegraphics[width=62mm]{fr10-\x .eps}\hspace{-23mm}\hspace{-2mm}
\includegraphics[width=62mm]{fr30-\x .eps}\hspace{-23mm}\hspace{-2mm}
\includegraphics[width=62mm]{fr70-\x .eps}\hspace{-23mm}\\
}
\caption{Capture fraction $f$ of the 5:2 inner resonance. Color and symbol codes are the same as in Figure 1.}
\end{center}
\end{figure*}

\bibliographystyle{mn2e}

\bibliography{ms}

\begin{thebibliography}{}

\bibitem[\protect\citeauthoryear{{Bailey} \& {Malhotra}}{{Bailey} \&
  {Malhotra}}{2009}]{BaileyMalhotra09}
{Bailey} B.~L.,  {Malhotra} R.,  2009, \icarus, 203, 155

\bibitem[\protect\citeauthoryear{{Borderies} \& {Goldreich}}{{Borderies} \&
  {Goldreich}}{1984}]{BorderiesGoldreich84}
{Borderies} N.,  {Goldreich} P.,  1984, Celestial Mechanics, 32, 127

\bibitem[\protect\citeauthoryear{{Engels} \& {Henrard}}{{Engels} \&
  {Henrard}}{1994}]{Henrard4}
{Engels} J.~R.,  {Henrard} J.,  1994, Celestial Mechanics and Dynamical
  Astronomy, 58, 215

\bibitem[\protect\citeauthoryear{{Fern{\'a}ndez}, {Gallardo} \&
  {Young}}{{Fern{\'a}ndez} et~al.}{2016}]{Fernandez16}
{Fern{\'a}ndez} J.~A.,  {Gallardo} T.,    {Young} J.~D.,  2016, \mnras, 461,
  3075

\bibitem[\protect\citeauthoryear{{Gronchi} \& {Milani}}{{Gronchi} \&
  {Milani}}{1999}]{GronchiMilani99}
{Gronchi} G.~F.,  {Milani} A.,  1999, \aap, 341, 928

\bibitem[\protect\citeauthoryear{{Henrard}}{{Henrard}}{1982}]{Henrard1}
{Henrard} J.,  1982, Celestial Mechanics, 27, 3

\bibitem[\protect\citeauthoryear{{Henrard} \& {de Vleeschauwer}}{{Henrard} \&
  {de Vleeschauwer}}{1988}]{Henrard2}
{Henrard} J.,  {de Vleeschauwer} A.,  1988, Celestial Mechanics, 43, 99

\bibitem[\protect\citeauthoryear{{Henrard} \& {Moons}}{{Henrard} \&
  {Moons}}{1992}]{Henrard3}
{Henrard} J.,  {Moons} M.,  1992, \icarus, 95, 244

\bibitem[\protect\citeauthoryear{{Katz}, {Dong} \& {Malhotra}}{{Katz}
  et~al.}{2011}]{Katzetal11}
{Katz} B.,  {Dong} S.,    {Malhotra} R.,  2011, Physical Review Letters, 107,
  181101

\bibitem[\protect\citeauthoryear{{Kozai}}{{Kozai}}{1962}]{Kozai62}
{Kozai} Y.,  1962, \aj, 67, 591

\bibitem[\protect\citeauthoryear{{Lemaitre}}{{Lemaitre}}{1984}]{Lemaitre84}
{Lemaitre} A.,  1984, Celestial Mechanics, 32, 109

\bibitem[\protect\citeauthoryear{{Lidov}}{{Lidov}}{1962}]{Lidov62}
{Lidov} M.~L.,  1962, \planss, 9, 719

\bibitem[\protect\citeauthoryear{{Lithwick} \& {Naoz}}{{Lithwick} \&
  {Naoz}}{2011}]{LithwickNaoz11}
{Lithwick} Y.,  {Naoz} S.,  2011, \apj, 742, 94

\bibitem[\protect\citeauthoryear{{Morais} \& {Giuppone}}{{Morais} \&
  {Giuppone}}{2012}]{MoraisGiuppone12}
{Morais} M.~H.~M.,  {Giuppone} C.~A.,  2012, \mnras, 424, 52

\bibitem[\protect\citeauthoryear{{Morais} \& {Namouni}}{{Morais} \&
  {Namouni}}{2013}]{MoraisNamouni13b}
{Morais} M.~H.~M.,  {Namouni} F.,  2013, \mnras, 436, L30

\bibitem[\protect\citeauthoryear{{Morais} \& {Namouni}}{{Morais} \&
  {Namouni}}{2016}]{MoraisNamouni16}
{Morais} M.~H.~M.,  {Namouni} F.,  2016, Celestial Mechanics and Dynamical
  Astronomy, 125, 91

\bibitem[\protect\citeauthoryear{{Murray} \& {Dermott}}{{Murray} \&
  {Dermott}}{1999}]{ssdbook}
{Murray} C.~D.,  {Dermott} S.~F.,  1999, {Solar system dynamics}.
{Cambridge University Press}

\bibitem[\protect\citeauthoryear{{Namouni} \& {Morais}}{{Namouni} \&
  {Morais}}{2015}]{NamouniMorais15}
{Namouni} F.,  {Morais} M.~H.~M.,  2015, \mnras, 446, 1998

\bibitem[\protect\citeauthoryear{{Poincar{\'e}}}{{Poincar{\'e}}}{1902}]{Poincare02}
{Poincar{\'e}} H.,  1902, Bulletin Astronomique, Serie I, 19, 289

\bibitem[\protect\citeauthoryear{{Quillen}}{{Quillen}}{2006}]{Quillen06}
{Quillen} A.~C.,  2006, \mnras, 365, 1367

\bibitem[\protect\citeauthoryear{{Volk} \& {Malhotra}}{{Volk} \&
  {Malhotra}}{2013}]{VolkMalhotra13}
{Volk} K.,  {Malhotra} R.,  2013, \icarus, 224, 66

\bibitem[\protect\citeauthoryear{{Wisdom}}{{Wisdom}}{1980}]{Wisdom80}
{Wisdom} J.,  1980, \aj, 85, 1122

\bibitem[\protect\citeauthoryear{{Wisdom}}{{Wisdom}}{1982}]{Wisdom82}
{Wisdom} J.,  1982, \aj, 87, 577

\bibitem[\protect\citeauthoryear{{Wisdom}}{{Wisdom}}{1983}]{Wisdom83}
{Wisdom} J.,  1983, \icarus, 56, 51

\end{thebibliography}

\end{document}